\documentclass[intlimits,twoside,a4paper]{article}

\usepackage[cp1251]{inputenc}
\usepackage[eqsecnum]{cmpj3}

\usepackage[sort&compress,numbers]{natbib}
\usepackage{xcolor}

\issue{2022}{25}{1}{13702}
\doinumber{10.5488/CMP.25.13702}
\title[Theoretical analysis of magnetic properties]%
{Theoretical analysis of magnetic properties and the magnetocaloric effect using the Blume--Capel model%
}

\author[S. Oliveira, R. H. M. Morais, J. P. Santos, F. C. S\'{a} Barreto]{S. Oliveira\orcid{0000-0002-8536-2144}\refaddr{label1,label2}\thanks{Corresponding author: \email{samuel.de.oliveira@ifmg.edu.br}.},
        R. H. M. Morais\orcid{0000-0001-8657-1805}\refaddr{label1},
				J. P. Santos\orcid{0000-0001-5357-7151}\refaddr{label1,label3},
				F. C. S\'{a} Barreto\refaddr{label4}}
\addresses{
\addr{label1} Department of Natural Sciences, Federal University of S\~{a}o Jo\~{a}o del-Rei (UFSJ), Brazil
\addr{label2} Department of Science and Languages, Federal Institute of Education, Science and Technology of Minas Gerais (IFMG), Brazil
\addr{label3} Department of Mathematics, Federal University of S\~{a}o Jo\~{a}o del-Rei (UFSJ), Brazil
\addr{label4} Emeritus Professor, Department of Physics, Federal University of Minas Gerais (UFMG), Brazil
}


\date{Received Septmeber 20, 2021, in final form December 06, 2021}

\begin{document}

\maketitle

\begin{abstract}
This work investigates the magnetic properties and the magnetocaloric effect in the spin-1 Blume--Capel model. The study was carried out using the mean-field theory from the Bogoliubov inequality to obtain the expressions of free energy, magnetization and entropy. The magnetocaloric effect was calculated from the variation of the entropy obtained by the mean-field theory. Due to the dependence on the external magnetic field and the anisotropy included in the model, the results for the magnetocaloric effect provided the system with first-order and continuous phase transitions. To ensure the results, the Maxwell relations were used in the intervals where the model presents continuous variations in magnetization and the Clausius--Clapeyron equation in the intervals where the model presents discontinuity in the magnetization. The methods and models for the analysis of a magnetic entropy change and first-order and continuous magnetic phase transitions, such as mean-field theory and the Blume--Capel model, are  useful tools in understanding the nature of the magnetocaloric effect and its physical relevance.%
\printkeywords
\end{abstract}

\section{Introduction}

The magnetocaloric effect (MCE) is defined as the change in temperature in magnetic solids due to the application of a variable magnetic field. This effect can be measured and/or calculated by analyzing the adiabatic variation in temperature $\Delta T_{\rm {ad}}(T,\Delta H)$ or an isothermal variation in magnetic entropy $\Delta S_{\rm {m}}(T,\Delta H)$ and the interaction of the material with a variable magnetic field~\cite{Pecharsky-1999}. First observed by Warburg~\cite{Warburg-1881}, the MCE was scientifically explained only years later by Weiss and Picard in their work on the phenomenon~\cite{Weiss-1918}. Currently, this phenomenon has attracted the attention of researchers and industry due to its high potential for application in the development of cooling or heating devices~\cite{Tishin-2016}.

Research on the properties of MCE and its applications has been growing both in the theoretical research~\cite{Oliveira2010,Galisova-2014,Hamad-2013,Franco-2018,Alzahrani-2020,Szalowski-2020,Karlova-2020,Umit-2021} and in the experimental research~\cite{Ram-2018,Mao-2020,Ghorai-2020,Sharmaa-2020,Zhanga-2020,Saidi-2021,Christian-2018}. From a theoretical point of view, statistical mechanics models are effective tools in the study of the thermodynamic and magnetic properties of materials. Some models analyze the MCE using different methods. Recently, the properties of MCE were studied by several theoretical methods, namely, mean-field theory (MFT)~\cite{Amaral-2010,Amaral-2007,Garcia-2020,Hsinia-2018,Lamouri-2018,Hsini-2018,Salha-2021,Amhoud-2021}, effective-field theory~(EFT)~\cite{Laghrissi-2016,Vatansever-2017,Yuksel-2018.1,Yuksel-2018.2,Jurcisinova-2019,Jurcisinova-2020}, Monte Carlo (MC) simulation~\cite{Masrour-2016,Masrour-2017,Elyacoubi-2018,Singh-2019,Amhoud-2020, Sun-2021} and first-principles calculations~\cite{Salmani-2018, Ennassiria-2018, Charif-2020}.

Regarding the models used in the study of magnetocaloric properties, what stands out mostly is the spin-1/2 Ising model~\cite{Mohylna-2019, Guerrero-2020, Semjam-2020, Amraoui-2020, Chandra-2021,Akinci-2021}. In addition to this, other models can be used in the study of magnetic and thermodynamic characteristics of magnetic materials, such as: Ashkin--Teller (AT) model~\cite{Jander-2015, Jander-2015.2, Akinci-2017, Jander-2018, Rafael-2019}, Blume--Emery--Griffiths (BEG) model~\cite {Ertas-2015, Jander-2016, Feraoun-2018, Thomaz-2016}, mixed-spin Ising model~\cite{Jabar-2016,Cardona-2017,Jander-2018.2,Jander-2019,Keskin-2020,Heydarinasab-2020} and Blume--Capel~(BC) model~\cite{Plascak-1993,Plascak-2003,Silva-2006,Jander-2015.3, Jander-2017, Albayrak-2019, Jascur-2020,Fadil-2021,Albayrak.2-2021}. The properties of MCE in the generalized spin-S model were studied by Y\"{u}ksel and collaborators~\cite{Yusuf-2021} using the MFT, and by Akinci and collaborators~\cite{Akinci-2018} who studied the MCE using the effective-field theory, where they concentrated their studies on continuous phase transitions. The Blume--Capel model is one of the most studied spin-S models in statistical mechanics, due to its broad theoretical interest and practical applications. It was introduced by Blume~\cite{Blume-1966} and independently by Capel~\cite{Capel-1966}, and has been applied in the description of many different physical situations. In this model, for the specific case of spin-1, the results indicate the existence of first-order and continuous phase transitions, with zero external field, between the ordered ferromagnetic ($\rm F$) phases of low temperature, to a paramagnetic phase~($\rm P$) high-temperature disorder occurring at a critical temperature which is dependent on a single-ion anisotropy $D$.

The objective of the present work is to study the magnetocaloric effect in the spin-1 Blume--Capel model, in the lattice with coordination number $z=6$, in order to analyze its properties as a function of the anisotropy that the model presents, highlighting the regions in which the model presents first-order phase transition. This study was carried out from the MFT starting from the variational principle of Gibbs--Bogoliubov~\cite{Bogoliubov-1947,Feynman-1955,Falk-1970} to obtain results for the thermodynamic quantities such as free energy, entropy and magnetization. With the MFT results, we analyzed the MCE by the variation of the entropy obtained using the Maxwell relations for the intervals in which the model presents continuous variations in magnetization and the Clausius--Clapeyron equation for the intervals that present discontinuity in the magnetization. With those expressions, a detailed analysis of the magnetic properties as a function of temperature, anisotropy, and an external field was obtained. With these results, we determine the magnetizations that present first-order phase transitions, changing the values of the external field $h\geqslant 0$, in order to analyze them and to obtain a relationship between critical temperatures and critical external fields in these transitions. In the end, a detailed study of the MCE was made for anisotropy values that present continuous and first-order phase transitions for different external field values. The MFT is used in this work as a tool suitable for a first insight into the problem. Although the MFT approach does not provide accurate results compared to numerical methods, we can say that this is a good approach for a first qualitative insight into the problem.

In section 2, we present the spin-1 Blume--Capel model based on the variational principle of Gibbs--Bogoliubov, in order to obtain results for thermodynamic quantities such as free energy, entropy and magnetization, as well as the Maxwell and Clausius--Clapeyron relations. In section 3, we analyze numerical results and diagrams obtained from the application of MFT in the Blume--Capel model in the lattice with coordination number $z=6$. In Section 4, we present the closing remarks.

\section{Model and formulation}

The purpose of this section is to provide a mathematical description for the spin-1 Blume--Capel model using the MFT, starting from the Gibbs--Bogoliubov variational principle, in order to obtain results for thermodynamic quantities such as free energy, entropy, and magnetization, as well as the Maxwell and Clausius--Clapeyron relations.

The Hamiltonian of the spin-1 Blume--Capel model is given by
\begin{eqnarray}
\label{eq:1}
\displaystyle H= - J \sum_{\left\langle i,j\right\rangle}^{Nz/2} S_i S_j - D\sum_{i=1}^{N} S_i^2 - h\sum_{i=1}^{N} S_i, 
\end{eqnarray}
\noindent where $J$ is the coupling between the spin variables, $D$ is a single-ion anisotropy, $\left\langle i,j\right\rangle$ denotes a pair of nearest-neighboring spins, each $S_i$ is restricted to $(-1, 0,+1)$, $N$ represents the number of sites in the lattice, $z$ is the coordination number and $h$ is an external field.

Bogoliubov's variational principle is based on the validity of inequality
\begin{eqnarray}
\label{eq:2}
\displaystyle G \leqslant \Phi \equiv G_0 + \left\langle H - H_0 \right\rangle, 
\end{eqnarray}
\noindent where $G$ and $H$ represent the free energy and the total Hamiltonian of a lattice statistical model, while~$G_0$ and $H_0$ represent the free energy and the trial Hamiltonian of a lattice statistical model simplified for which the relevant calculations can be performed exactly (the symbol $\left\langle ...\right\rangle$ denotes the average by the canonical ensemble within the simplified model defined by the Hamiltonian $H_0$). The term $\Phi$ is the variational free energy, which gives an upper bound on the true free energy $G$.

The trial Hamiltonian of the spin-1 Blume--Capel model must include in the MFT a variational parameter $\gamma$:
\begin{eqnarray}
\label{eq:3}
\displaystyle H_0 = - \gamma \sum_{k=1}^{N} S_k - D \sum_{k=1}^{N} S_k^2.
\end{eqnarray}
To obtain the above variational parameter, we must minimize the free energy as presented more in detail in the references~\cite{Plascak-1993,Jander2019.2,Jander-2017}. After the calculations, we obtain expressions for free energy and magnetization, given by
\begin{eqnarray}
\label{eq:4}
\displaystyle 
G = -N k_{\rm B} T \ln \left\{ 1 + 2 \re^{\beta D} \cosh \left[\beta(zJm+h) \right] \right\} + \frac{Nz}{2} Jm^2,
\end{eqnarray}
\noindent and
\begin{eqnarray}
\label{eq:5}
\displaystyle 
m = \frac{2\re^{\beta D} \sinh[\beta(zJm + h)]}{1 + 2\re^{\beta D} \cosh [\beta(zJm + h)]}. 
\end{eqnarray}
\noindent Here, $\beta \equiv 1/k_{\rm B} T$, $k_{\rm B}$ is the Boltzmann constant and $T$ is the temperature. From equations~(\ref{eq:4}) and~(\ref{eq:5}), we obtain the entropy by the expression,
\begin{eqnarray}
\label{eq:7}
S=-\frac{\partial G}{\partial T}.
\end{eqnarray}

The variation of entropy, $\Delta S_{m}(T,\Delta h)$, is obtained by taking the difference between the entropy in the final and initial states of an isothermal process, i.e.,
\begin{eqnarray}
\label{eq:8}
\Delta S_{m}(T,\Delta h) = S_{m_2}(T, h_2) - S_{m_1}(T, h_1).
\end{eqnarray}

An alternative way to obtain the MCE is through the Maxwell relations, from equations~(\ref{eq:5}) and~(\ref{eq:7})~\cite{Tishin-2016}, given by the integral
\begin{eqnarray}
\label{eq:9}
\displaystyle \Delta S_{m}(T, \Delta h) = \int^{h_2}_{h_1} \left[\frac{\partial m(T,h)}{\partial T}\right]_h \rd h,
\end{eqnarray}

\noindent where $\Delta S_{m}$ is obtained for the values of $T$ for which only continuous phase transitions occur.

For systems with first-order phase transitions, we use equation~(\ref{eq:9}) with a contribution present in the entropy variation due to discontinuity in magnetization, in order to obtain $\Delta S_{m}(T, \Delta h)$ which takes the following form,
\begin{eqnarray}
\label{eq:10}
\displaystyle \Delta S_{m}^{(of)}(T, \Delta h) \cong \int^{h_{\rm{C}}-\delta h}_{h_1} \left[\frac{\partial m(T,h)}{\partial T}\right]_h \rd h +\int^{h_2}_{h_{\rm{C}}+\delta h} \left[\frac{\partial m(T,h)}{\partial T}\right]_h \rd h + \delta S,
\end{eqnarray}

\noindent  where the variations represented in the intervals of the integrals tend to zero ($\delta h \rightarrow 0$) and $(of)$ denotes the first-order phase transitions under the critical conditions $T_{\rm{C}}$ and $h_{\rm{C}}$. The expression given by equation~(\ref{eq:10}) has three terms, the first and the second of which make it possible to determine the MCE~($\Delta S_{m}$) by the magnetization derivative, in the magnetic field intervals where variations in magnetizations are continuous. The third term allows obtaining the contribution of the MCE due to discontinuity, due to the first-order phase transitions, in magnetization and will be obtained by the Clausius--Clapeyron equation, given by,
\begin{eqnarray}
\label{eq:11}
\delta S = \left(\frac{\rd h}{\rd T}\right) \delta m(T_{\rm{C}}, h_{\rm{C}}),
\end{eqnarray}

\noindent where the magnetization variation ($\delta m$) around $T_{\rm{C}}$ is determined by $\delta m(T_{\rm{C}}, h_{\lambda})=m_{h_{\rm{C}}-\delta h}-m_{h_{\rm{C}}+ \delta h}$.


\section{Numerical results and diagrams}

In this section we present numerical results and diagrams obtained from the application of the MFT to the spin-1 Blume--Capel model on the lattice with coordination number $z=6$. With the equations~(\ref{eq:4}--\ref{eq:11}) presented in the last section, we obtain the phase diagrams ($T{-}D$) and ($T{-}h$), the diagrams for magnetization ($m$) as a function of temperature ($T$), of the free energy ($G$) as a function of magnetization, of entropy ($S$) as a function of temperature and the diagram for the ($-\Delta S_{m}$) as a function of temperature. All MCE results were obtained by the variation of the entropy given by equation~(\ref{eq:8}), and by the Maxwell relations given by equations~(\ref{eq:9}--\ref{eq:11}), and the results were identical. For simplicity, we consider $k_{\rm B} = 1$ and $J = 1$.

Before starting the analysis of the MCE properties, we discuss the characteristics of the phase transition diagrams of the model ($T{-}D$ and $T{-}h$) considering some external magnetic field values ($h \geqslant 0 $). The phases and regions in which the model presents are classified as follows: 

Considering the null external field ($h=0$), we define: 

$(i)$ The ferromagnetic phase ($\rm F$), with $m \neq 0$.

$(ii)$ The paramagnetic phase ($\rm P$), with $m = 0$.

$(iii)$ The dense ferromagnetic region ($\rm{d_F}$), in which the phase $\rm F$ is stable and $\rm P$ is metastable.

$(iv)$ The dense paramagnetic region ($\rm{d_P}$), in which the phase $\rm P$ is stable and $\rm F$ is metastable. 

Considering the external field greater than zero ($h > 0$), we define: 

$(v)$ The ordered phase (${\rm {OP}}_1$) with $m \neq 0$.

$(vi)$ The ordered phase (${\rm {OP}}_2$) with $m \neq 0$.

$(vii)$ The dense region (${\rm d}_{{\rm {OP}}_1}$) in which the phase ${\rm {OP}}_1$ is stable and ${\rm {OP}}_2$ is metastable.

$(viii)$ The dense region (${\rm d}_{{\rm {OP}}_2}$) in which the phase ${\rm {OP}}_2$ is stable and ${\rm {OP}}_1$ is metastable.

\begin{figure}[h!]
	\centering
\begin{minipage}{0.49\linewidth}
			\includegraphics[width=0.9\textwidth]{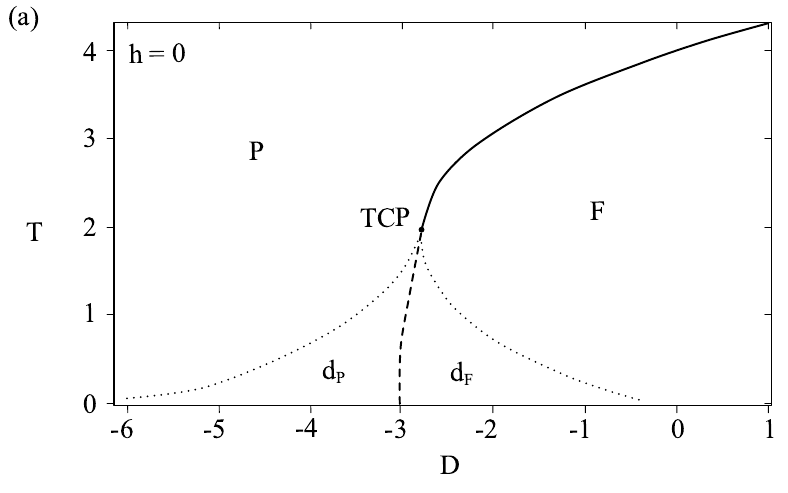}
			\end{minipage} \hfill
\begin{minipage}{0.49\linewidth}
			\includegraphics[width=0.9\textwidth]{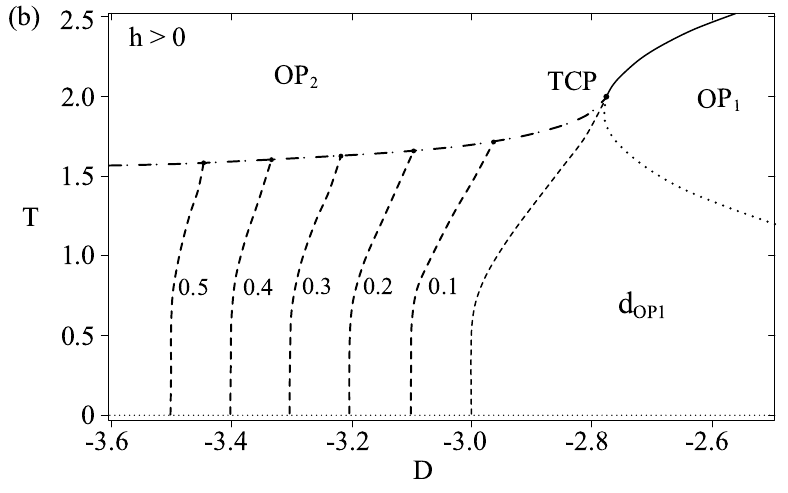}
			\end{minipage} \hfill
\begin{minipage}{0.49\linewidth}
			\includegraphics[width=0.9\textwidth]{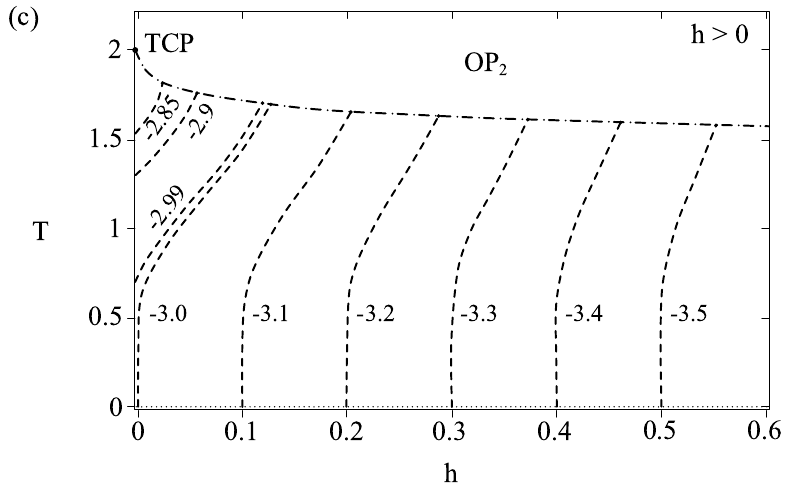}
			\end{minipage} \hfill
\begin{minipage}{0.49\linewidth}
			\includegraphics[width=0.9\textwidth]{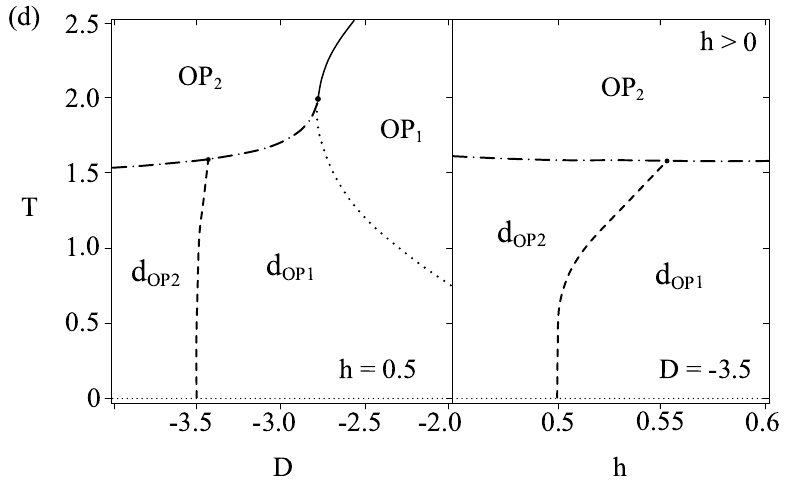}
			\end{minipage} \hfill	
  		\caption{Phase diagrams obtained by MFT for the spin-1 Blume--Capel model with coordination number $z=6$. (a) Phase diagram ($T{-}D$) showing dashed line of first-order phase transitions from steady state to null field $h = 0$. (b) Phase diagram ($T{-}D$) for some values of the external magnetic field $h > 0$. The numbers that follow each curve represent some values of $h$. (c) Phase diagram ($T{-}h$) explaining the coexistence line of the regions (${\rm d}_{{\rm {OP}}_1}$) and (${\rm d}_{{\rm {OP}}_2}$) for some values of $D$. The numbers that follow each curve represent the values of $D$. (d) Phase diagram ($T{-}D$) and ($T{-}h$) for the fixed values $h=0.5$ and $D=-3.5$, respectively, highlighting the regions ${\rm d}_{{\rm {OP}}_1}$ and ${\rm d}_{{\rm {OP}}_2}$.}
		\label{fig:1}
\end{figure}

The phases defined above are shown in figure~\ref{fig:1} where the solid lines represent continuous phase transitions, the dashed lines represent the first-order phase transitions, the critical lines dashed and dotted represent the limit for the first-order phase transitions to different external field values, and lastly the dotted lines limit the dense region.

Considering a null external field, we use the same methodology applied in reference~\cite{Jander-2017}, and we present the results for the phase diagram $T{-}D$ in figure~\ref{fig:1}~(a), whose obtained phases are $\rm F$ and $\rm P$, and also show the named regions $\rm{d_F}$ and $\rm{d_P}$. Still, in the figure it is possible to verify that for the values of $D = 0$, $D \rightarrow \infty$, and $D = -2.772$, the temperature values are approximately $T = 4.0, 6.0$, and $2.0$, respectively. The latter is the temperature of the tricritical point  $(\rm{TCP})$.

For non-zero external field ($h>0$), we present three different phase diagrams in figures~\ref{fig:1}~(b), (c) and~(d), whose obtained phases are $\rm{OP}_1$ and $\rm{OP}_2$, as well the regions ${\rm d}_{{\rm {OP}}_1}$ and ${\rm d}_{{\rm {OP}}_2}$. In figure~\ref{fig:1}~(b), we present the phase diagram ($T{-}D$) for some external magnetic field values between $h =0.1$ and $h=0.5$, where it is possible to observe that as we increase the external magnetic field, the first-order phase transition line between regions ${\rm d}_{{\rm {OP}}_1}$ and ${\rm d}_{{\rm {OP}}_2}$ shifts in the negative direction of $D$. These lines of first-order phase transitions (dashed lines) were obtained from equations~(\ref{eq:4}) and (\ref{eq:5}), from a comparison of the Gibbs free energies. The energies were compared considering the condition $(G_{\rm F} = G_{\rm P})$, where $G_{\rm F}$ represents the free energy of the ferromagnetic phase and $G_{\rm P}$ represents the free energy of the paramagnetic phase, and also $(G_{\rm{OP}_1} = G_{\rm{OP}_2})$, where $G_{\rm{OP}_1}$ represents the free energy of the ordered phase $\rm{OP}_1$ and $G_{\rm{OP}_2}$ represents the free energy of the ordered phase $\rm{OP}_2$, which converges to $D+h=-z/2$ with ($T \rightarrow 0$). The critical line (dashed and dotted) starts at the tricritical point $\rm{TCP} = (-2.772, 2.000)$ and tends to the temperature $T=1.5$ with $D\rightarrow -\infty$. In figure~\ref{fig:1}~(c), we present the phase diagram ($T{-}h$) for some fixed values of $D$, ranging between $D =-2.85$ and $D=-3.5 $, where it is verified that an increase in the anisotropy leads to an increase in the magnetic field $h$ for which the first-order phase transition line between the regions ${\rm d}_{{\rm {OP}}_1}$ and ${\rm d}_{{\rm {OP}}_2}$ occurs. It is also possible to observe that as the value of $D$ gets closer to the critical value $(-2.772)$, the coexistence curve gets closer to the vertical axis indicating a decrease of the region ${\rm d}_{{\rm {OP}}_2}$. To highlight the regions ${\rm d}_{{\rm {OP}}_1}$ and ${\rm d}_{{\rm {OP}}_2}$ we present in figure~\ref{fig:1}~(d) the phase diagrams ($T{-}D$) for the fixed value of $h = 0.5$, as well as ($T{-}h$) for the fixed value of $D = -3.5$.

\begin{figure}[h!]
	\centering
\begin{minipage}{0.49\linewidth}
			\includegraphics[width=0.9\textwidth]{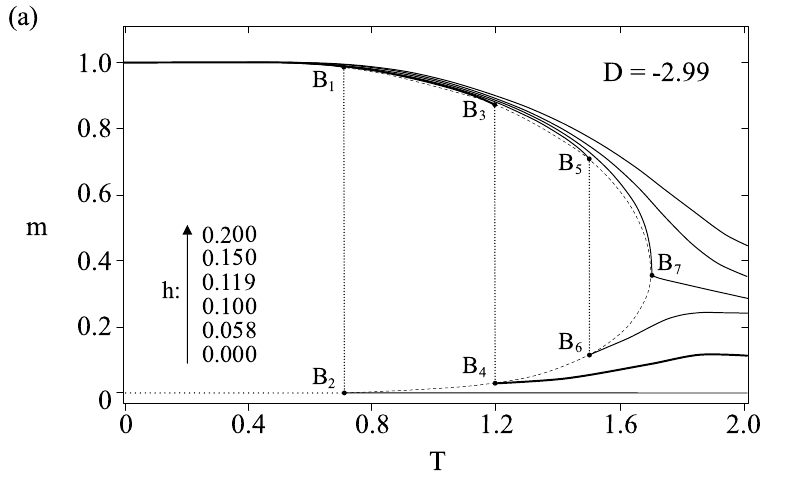}
			\end{minipage} \hfill
\begin{minipage}{0.49\linewidth}
			\includegraphics[width=0.9\textwidth]{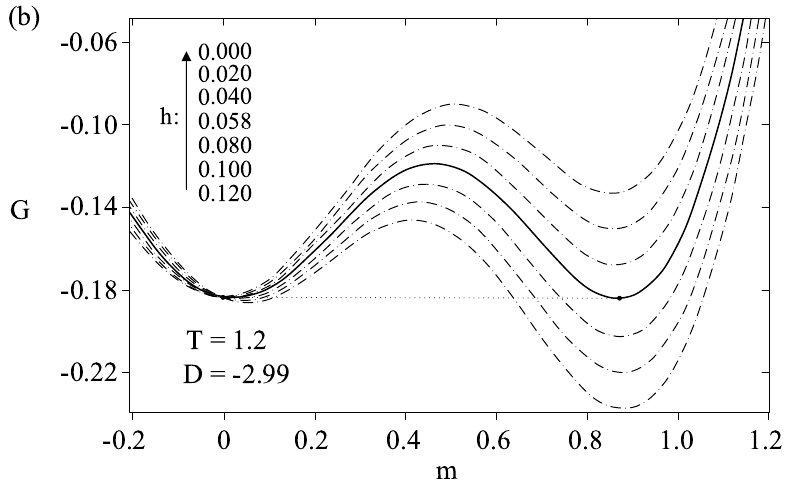}
			\end{minipage} \hfill
\begin{minipage}{0.49\linewidth}
			\includegraphics[width=0.9\textwidth]{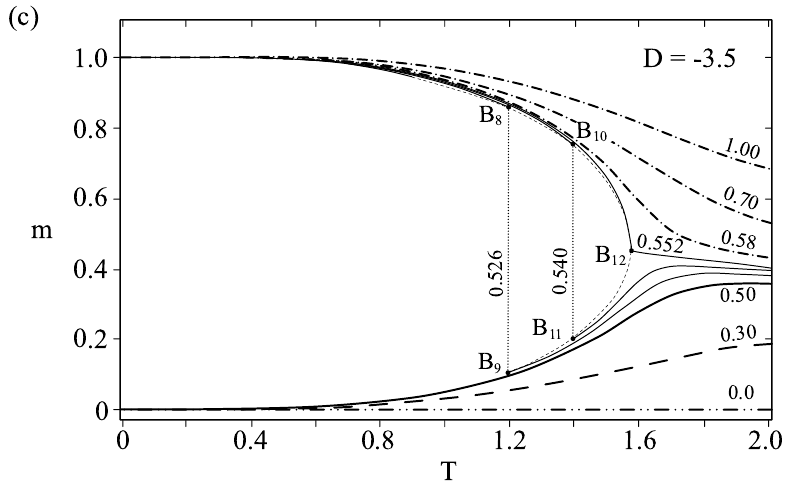}
			\end{minipage} \hfill
\begin{minipage}{0.49\linewidth}
			\includegraphics[width=0.9\textwidth]{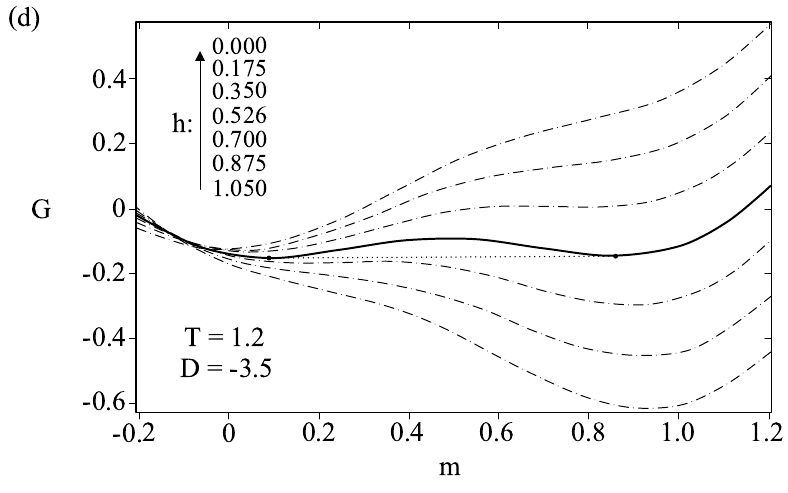}
			\end{minipage} \hfill	
	  	\caption{(a) Magnetization as a function of temperature for the value $D=-2.99$ with external field values $h = \{0, 0.0589, 0.1, 0.1193, 0.15, 0.2\} $. (b) Free energy as a function of magnetization for the value $D=-2.99$ with external field values $h = \{0, 0.02, 0.04, 0.058, 0.08, 0.1, 0.12\}$. The data obtained are for $T=1.2$. (c) Magnetization as a function of temperature for the value $D=-3.5$ with external field values $h = \{0, 0.3, 0.5, 0.526, 0.54, 0.552, 0.58 , 0.7\}$. The numbers that follow each curve represent the values of $h$ at which it was calculated in the model. (d) Free energy as a function of magnetization for the value $D=-3.5$ with external field values $h = \{ 0, 0.175, 0.350, 0.526, 0.7, 0.875, 1.05 \}$. The data obtained are for $T = 1.2$.}
		\label{fig:2}
\end{figure}

To analyze the behavior of the phases that define the phase diagrams in figure~\ref{fig:1}, we present in figure~\ref{fig:2} the magnetization diagrams as a function of temperature, and the free energy diagrams as a function of magnetization for the values of $D = -2.99$ and $D = -3.5$. Observing figure~\ref{fig:2}~(a), it is possible to verify that for $D = -2.99$ and the null external field $h = 0$, the system undergoes a first-order phase transition, represented by vertical line $B_1 B_2$. The other two vertical lines, represented by $B_3 B_4$ and $B_5 B_6$, correspond to the first-order phase transition for the external field values $h = 0.058$ and $h = 0.100$, respectively. Still, with respect to this diagram, it can be observed that from the application of an external field $h = 0.119$, the coexistence line associated with the first-order phase transition ceases to exist and the system evolves to continuous phase transitions, the behavior confirmed from the phase diagram shown in figure~\ref{fig:1}~(c), that is, the system passes to phase $\rm{OP}_2$ at temperatures higher than that shown in point the~$B_7$ in figure~\ref{fig:2}~(a). Still, in figure~\ref{fig:2}~(a) it is possible to verify that the dashed line connecting the points $B_1$, $B_3$, $B_5$, $B_7$, $B_6$, $B_4$, and $B_2$ limits the region where the continuous phase transitions occur. In figure~\ref{fig:2}~(b), we have a representation of the free energy for the fixed temperature $T = 1.2$ and anisotropy $D=-2.99$, where the system undergoes first-order phase transitions with external field  $h=0.058$, where for the external field range $h < 0.058$, the stable state is phase $\rm{OP}_1$, while for the range $h > 0.058$, the stable state is phase $\rm{OP}_2$.

In figure~\ref{fig:2}~(c), we present a magnetization diagram for $D = -3.5$, and it is possible to observe that the first-order phase transitions only happen with the application of an external field $h > 0.5$, confirmed by the phase diagram shown in figure~\ref{fig:1}~(d), and shown by the vertical line $B_8 B_9$, where the transition occurs with the application of an external field $h = 0.526$. The other vertical line, represented by $B_{10} B_{11}$ corresponds to the first-order phase transition for the external field $h = 0.540$, while the dashed line connecting points $B_8$, $B_{10}$, $B_{12}$, $B_{11}$, and $B_9$ limits the region where continuous phase transitions occur. For $h < 0.5$, the phase $\rm{OP}_2$ is stable, while the phase $\rm{OP}_1$ is unstable, as obtained by the results of the external field $h = \{ 0, 0.3, 0.5 \}$. We also show that by applying an external field $h = 0.552$, the coexistence line associated with the first-order phase transition ceases to exist and the system evolves to continuous transitions, which can be seen in the phase diagram shown in figure~\ref{fig:1}~(d) and by the point $B_{12}$ in figure~\ref{fig:2}~(c). Furthermore, in figure~\ref{fig:2}~(d), we present the free energy for the fixed temperature $T = 1.2$, and anisotropy $D=-3.5$, where the system undergoes first-order phase transitions with external field $h=0.526$, where for the external field range $h < 0.526$, the stable state is phase $\rm{OP}_1$, while for the range $h > 0.526$, the stable state is phase $\rm{OP}_2$ .

\begin{figure}[h!]
	\centering
\begin{minipage}{0.49\linewidth}
			\includegraphics[width=1\textwidth]{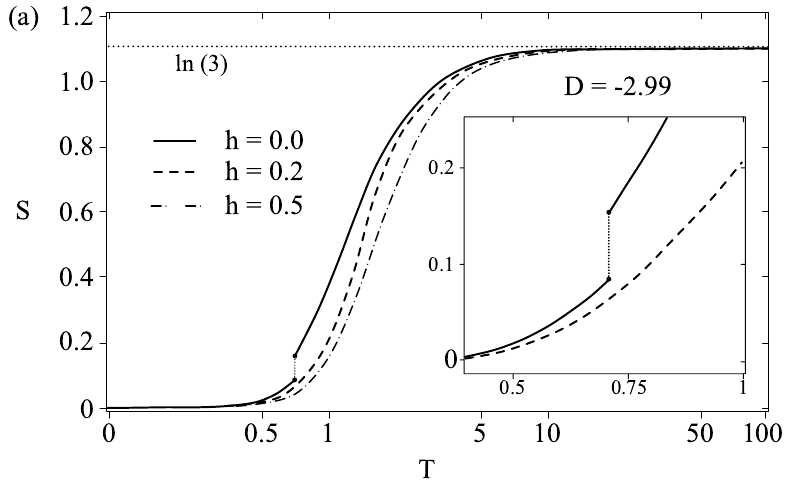}
			\end{minipage} \hfill	
\begin{minipage}{0.49\linewidth}
			\includegraphics[width=1\textwidth]{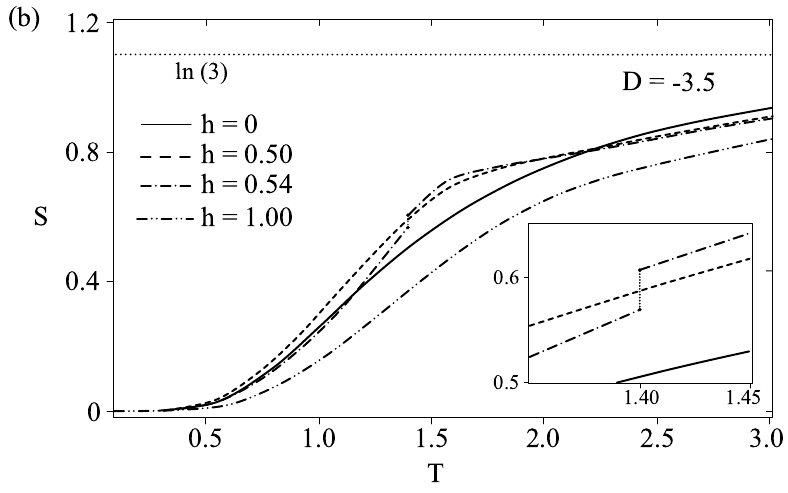}
			\end{minipage} \hfill	
		  \caption{(a) Diagram for entropy as a function of temperature for three values of the applied magnetic field $h = \{0, 0.2, 0.5 \}$, using parameter $D = -2.99$. In the inner region of the diagram it is possible to observe the first-order phase transition to null field, at approximately $T = 0.71$. (b) Diagram for entropy as a function of temperature for four values of the applied magnetic field $h = \{ 0, 0.5, 0.54, 1.0 \}$, using parameter $D = -3.5$ . In the inner region of the diagram it is possible to observe the first-order phase transition to field $h=0.54$ at temperature $T=1.4$.}
		\label{fig:3}
\end{figure}

Before presenting the MCE results, we study the behavior of entropy as a function of temperature, for some external field and anisotropy values $D=\{-2.99, -3.5\}$, and the results can be seen in figure~\ref{fig:3}. In figure~\ref{fig:3}~(a) we present the entropy as a function of temperature for three values of the applied magnetic field $h = \{ 0, 0.2, 0.5 \}$, in which it is possible to verify that the entropy with a null field displays a discontinuity at approximately $T = 0.71$, confirming the result presented in figure~\ref{fig:2}~(a) in the line $B_1 B_2$. Analyzing the entropy as a function of temperature, now for the parameter $D = -3.5$, we present in figure~\ref{fig:3}~(b) the result for four different values of the applied magnetic field $h = \{0 , 0.5, 0.54, 1.0 \}$. As discussed previously, for this parameter the first-order phase transition between the regions ${\rm d}_{{\rm {OP}}_1}$ and ${\rm d}_{{\rm {OP}}_2}$ at low temperatures occurs with the application of a magnetic field in the interval $0.5 < h < 0.552$. Thus, we observe in this figure that the entropy for $h = 0.54$ shows a first-order phase transition that appears at approximately $T = 1.4$, confirmed by the phase diagram in figure~\ref{fig:1}~(d), and can also be seen in figure~\ref{fig:2}~(c), through the line $B_{10} B_{11}$. For external field $h = 0$, the stable phase is $\rm P$, for $h=0.5$ the system is in the region ${\rm d}_{{\rm {OP}}_2}$ for low temperatures showing continuous phase transitions for phase $\rm{OP}_2$, while for external field $h = 1.0$, the system is in the region ${\rm d}_{{\rm {OP}}_1}$ for low temperatures showing continuous phase transitions for the phase $\rm{OP}_2$.

\begin{figure}[h!]
	\centering
\begin{minipage}{0.49\linewidth}
		\includegraphics[width=0.9\textwidth]{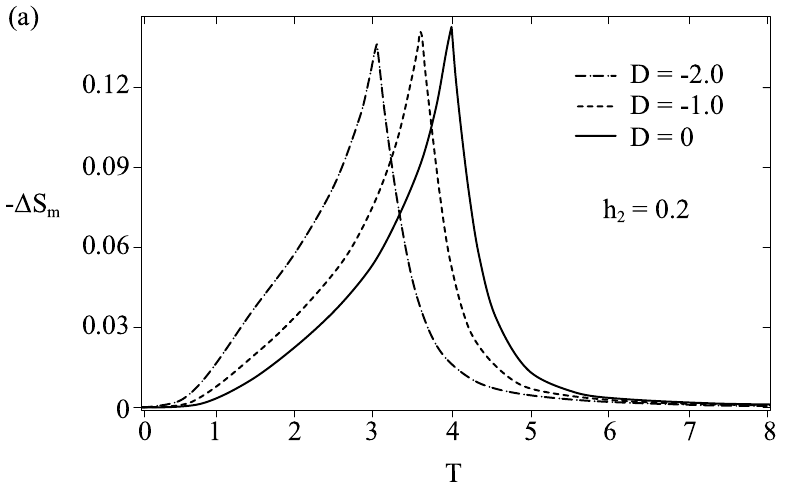}
		\end{minipage} \hfill
\begin{minipage}{0.49\linewidth}
		\includegraphics[width=0.9\textwidth]{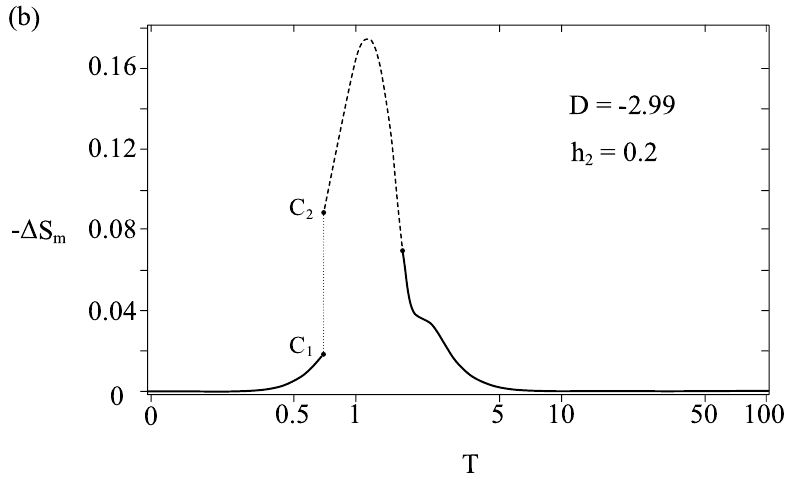}
		\end{minipage} \hfill	
	  \caption{(a) Variation of the magnetic entropy for the values of $D = 0, -1$ and $-2$, with a maximum applied field of $h = 0.2$. (b) Variation of magnetic entropy for $D = -2.99$, with a maximum applied field of $h = 0.2$. The line $C_1 C_2$ represents the first-order phase transition that occurs approximately at $T=0.71$.}
		\label{fig:4}
\end{figure}

Finally, we present the results obtained for the MCE from equations~(\ref{eq:8})--(\ref{eq:11}), coupled with equations~(\ref{eq:5}) and (\ref{eq:7}). In our study, the MCE was obtained by equation~(\ref{eq:8}) and also by equations~(\ref{eq:9})--(\ref{eq:11}), the same results were determined, which was already predicted. To start the study, we present in figure~\ref{fig:4}~(a) the dependence of $-\Delta S_{m}$ on the temperature for the anisotropy values $D = \{-2, -1, 0 \}$ and with a maximum applied field of $h_2 = 0.2$, while $h_1 = 0$ throughout the whole study. With the results obtained we observed that the change in magnetic entropy gradually decreases both below and above the magnetic ordering temperature with a decrease of the anisotropy. As is already known in the literature~\cite{Oliveira2010,Yuksel-2018.2,Akinci-2018,Hcini2019,Wang2020,Hamad2020}, the MCE in models that present continuous phase transitions, show a behavior with $-\Delta S_{m}$ tending to zero with $T \rightarrow 0$, an increasing trend towards the critical temperature, peaking and declining more sharply, and tending to zero with $T \rightarrow \infty$. This behavior was discussed in the work by Franco and collaborators~\cite{Franco2006}, where the magnetic entropy change curves for different applied fields collapse into a single curve for materials with continuous phase transitions. It is important to report that for compounds which undergo a first-order phase transition, this universal curve is not valid. Thus, this result~\cite{Franco2006} suggests that the universal curve can be used as an additional criterion to distinguish the order of the phase transition.

To analyze the main objective of this work, which is the MCE, in models that present first-order phase transitions, we plot in figure~\ref{fig:4}~(b) the $-\Delta S_{m}$ with respect to temperature for the value of $D = -2.99$, with a maximum applied field of $h_2 = 0.2$. In the diagram we can see that with $T \rightarrow 0$ we have $(-\Delta S_{m}) \rightarrow 0$. Increasing the value of $T$, the curve presents an increasing trend in a region that presents only a continuous variation of the entropy (solid line). At the critical temperature $T_{\rm{C}} = 0.71$, a discontinuity occurs due to the first order phase transition ($C_1 C_2 $), which can be confirmed in figure~\ref{fig:2}~(a). From point $C_2$, the model presents a region with first order phase transitions (dashed line), passing to a region with only continuous variations (continuous line) at approximately $T = 1.7$, tending to zero with $T \rightarrow \infty$.

\begin{figure}[h!]
	\centering
\begin{minipage}{0.47\linewidth}
			\includegraphics[width=0.9\textwidth]{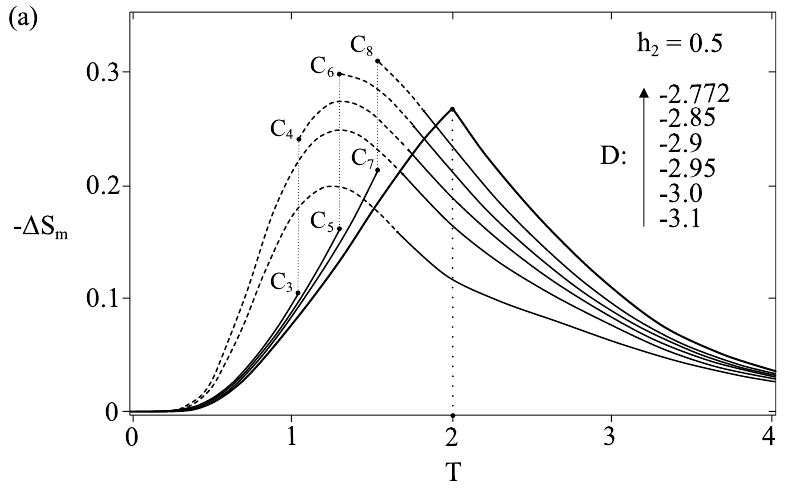}
			\end{minipage} \hfill
\begin{minipage}{0.47\linewidth}
			\includegraphics[width=0.9\textwidth]{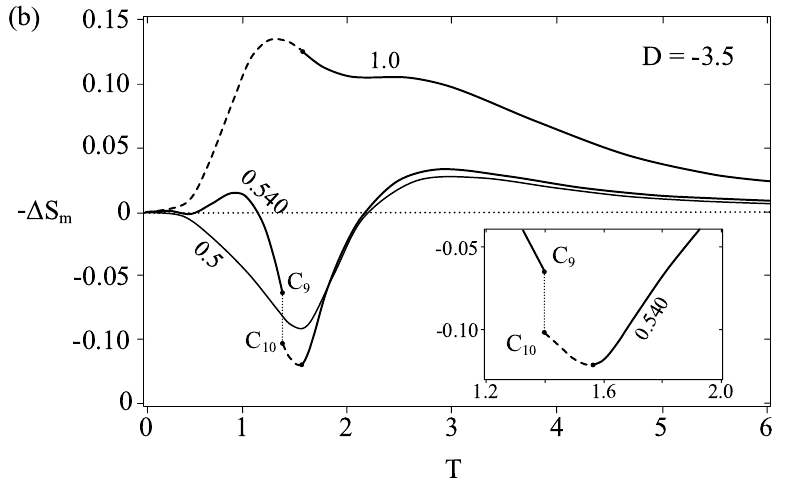}
			\end{minipage} \hfill	
		\caption{(a) Variation of magnetic entropy for the values of $D = \{-2.772, -2.85, -2.9, -3.0, -3.1\}$, with the maximum
applied field of $h_2 = 0.5$. (b) Variation of magnetic entropy for $D = -3.5$, and magnetic field intervals at $\Delta h = 0.5$, $\Delta h = 0.540$ and $\Delta h = 1.0$. The line $C_9 C_{10}$ represents the first-order phase transition that occurs approximately at $T = 1.4$. The numbers that follow each curve represent the values of $\Delta h$ that was calculated in the model.}
		\label{fig:5}
\end{figure}

In figure~\ref{fig:5}~(a) we present the trend of the ($-\Delta S_{m}$) in relation to the temperature for some values of~$D$ between the point TCP ($D=-2.77$) and $D=-3.1$, with a maximum applied field of $h_2 = 0.5$. For a value of $D=-2.77$, the model still does not show first-order phase transition, so ($-\Delta S_{m}$) behaves like the universal curve suggested in the work of Franco et al.~\cite{Franco2006}, where the MCE is maximum at the critical temperature. For anisotropy values $D < -2.77$ and the proposed magnetic field variation, the model presents first-order phase transitions up to $D=-3.1$. In the regions where first-order phase transitions occur, MCE is represented by dashed lines and in the regions where only continuous variations in magnetization occur, MCE is represented by solid lines. We can observe that for anisotropy values below the TCP point ($D<-2.77$), the MCE values at the points $C_iC_j$, of the vertical lines of discontinuity gradually lose intensity and reduce to zero until touching $\Delta S_{m}=0$ in $D=-3.0$. Below $D=-3.0$, the dashed line, obtained by the first-order phase transitions, continues to reduce the intensity until $D=-3.5$ which does not show first-order phase transitions for $h<0.5$ and this can be seen in figure~\ref{fig:5}~(b). Fixing at $D=-3.5$, figure~\ref{fig:5}~(b) shows the trends of the MCE when intensifying the variation of the external field $\Delta h$. For $\Delta h = 0.5$, the model does not present first-order phase transitions and the MCE is obtained only with continuous variations in magnetization. For $\Delta h = 0.540$, we identified, as predicted, a discontinuity in the magnetocaloric potential at approximately $T = 1.4$ ($C_9C_{10}$), at which point the stable phase passes from the region ${\rm d}_{{\rm {OP}}_1}$ to ${\rm d}_{{\rm {OP}}_2}$. Then, the system passes from the ${\rm d}_{{\rm {OP}}_2}$ to the $\rm{OP}_2$ phase through a continuous phase transition. For $\Delta h = 1.0$, the MCE presents the same behavior as shown in figure~\ref{fig:5}~(a) for $D= -3.0$ and $-3.1$.


\section{Conclusions}

In this work, the application of the MFT allowed us to analyze the magnetic and magnetocaloric properties using the spin-1 Blume--Capel model in the lattice with coordination number $z=6$, having, as variation, the anisotropy and external field quantities near critical temperatures. According to our results, depending on the value of the anisotropy, the system can exhibit a continuous or first-order phase transition behavior. We confirm that the MCE peaks obtained for anisotropy values greater than $D = -2.772$ is in agreement with the unique universal behavior found in materials where the phase transition is continuous, in contrast to the behavior of the MCE in the first-order phase transitions. The difference between the entropy curves change in the continuous and first-order phase transitions is clear, as shown in figure~\ref{fig:5}~(a), where we quantify the data dispersion in the ordered region. The application of MFT together with the use of the spin-1 Blume--Capel model allowed us to determine the magnetic and magnetocaloric properties, in which we show that these results can serve as an important tool to analyze and describe the MCE phenomena. The results found from the variation of entropy obtained by equation~(\ref{eq:8}), were compatible with the use of Maxwell relations equations~(\ref{eq:9}--\ref{eq:11}) where the model presented continuous variations in the magnetization as well as the use of the Clausius--Clapeyron equation in the intervals where the magnetization presented discontinuity. With this comparison of results it was shown that in the applications of the MFT, one can obtain explicit expressions for entropy and use it efficiently for the MCE. Although in general the MFT approach does not provide accurate results compared to other numerical methods, it provides the first qualitative view of the problem. In the future we may be looking into the numerical calculations of the system.

\section*{Acknowledgements}

J.~P.~Santos would like to thank the National Council for Scientific and Technological Development --- CNPq (No.~306404/2019-2) for financial support. R.~H.~M.~Morais acknowledges financial support in part by the Higher Education Personnel Improvement Coordination --- Brazil (CAPES) --- Finance Code (No.~88887.501260/2020-00).


\ukrainianpart

\title{Теоретичний аналіз магнітних властивостей і магнітокалоричного ефекту в моделі Блюма--Капеля}

\author{С. Олівейра\orcid{0000-0002-8536-2144}\refaddr{label1,label2},
	Р. Х. M. Морайш\orcid{0000-0001-8657-1805}\refaddr{label1},
	Ж. П. Сантуш\orcid{0000-0001-5357-7151}\refaddr{label1,label3},
	Ф. С. Са Баррето\refaddr{label4}}
\addresses{
	\addr{label1} Факультет природничних наук, Федеральний університет Сан-Жоао дель Рей, Бразилія
	\addr{label2} Факультет науки і мов, Федеральний інститут освіти, науки і технології Мінас-Жерайса, Бразилія
	\addr{label3} Факультет математики, Федеральний університет Сан-Жоао дель Рей, Бразилія
	\addr{label4} Почесний професор, факультет фізики, Федеральний університет Мінас-Жерайса, Бразилія
}

\makeukrtitle

\begin{abstract}
У цій роботі досліджуються магнітні властивості та магнітокалоричний ефект у спін-1 моделі Блюма--Капеля.
Обчислення проведено з використанням теорії середнього поля та нерівності Боголюбова; отримано вирази для вільної енергії, намагніченості та ентропії.
Магнітокалорічний ефект розраховано методом варіації ентропії, отриманої з теорії середнього поля.
Оскільки в моделі враховано зовнішнє маг\-нітне поле та анізотропію, з результатів для магнітокалоричного ефекту знайдені фазові переходи в системі (неперервні та першого роду).  В областях, де в системі спостерігається неперервна або стрибкоподібна поведінка намагніченості, використовуються співвідношення Максвелла або Клаузіуса--Клапейрона, відповідно. Методи та моделі, використані для аналізу змін у ентропії та досліджуваних магнітних фазових переходів, а саме, теорія середнього поля та модель  Блюма--Капеля, є корисними інструментами для покращення розуміння природи магнітокалоричного ефекту та його фізичної значущості.
	\keywords магнітокалоричний ефект, теорія середнього поля, ізотермічна зміна ентропії, фазовий перехід, модель Блюма--Капеля
\end{abstract}

\lastpage
\end{document}